\documentclass[10pt,a4paper]{article}
\usepackage[a4paper,margin=25mm]{geometry}
\usepackage[T1]{fontenc}
\usepackage{lmodern}           
\usepackage{graphicx}
\usepackage{longtable}
\usepackage{url}
\usepackage{verbatim}          
\usepackage{pgfplots}
\pgfplotsset{compat=1.18}
\usepackage{xcolor}
\usepackage{authblk}           
\usepackage{hyperref}          

\providecommand{\email}[1]{\href{mailto:#1}{#1}}

\title{Conversational Agents for Building Energy Efficiency -- Advising Housing Cooperatives in Stockholm on Reducing Energy Consumption}

\author[1,2]{Shadaab Ghani}
\author[1]{Anne Håkansson}
\author[2]{Oleksii Pasichnyi}
\author[2]{Hossein Shahrokni}

\affil[1]{Department of Computer Science, KTH Royal Institute of Technology, Stockholm, Sweden}
\affil[2]{Department of Sustainable Development, Environmental Science and Engineering (SEED), KTH Royal Institute of Technology, Stockholm, Sweden}
\affil[ ]{\email{shadaab@kth.se}}

\date{} 

\begin{document}
\maketitle

\begin{abstract}
Housing cooperative is a common type of multifamily building ownership in Sweden. Although this ownership structure grants decision-making autonomy, it places a burden of responsibility on cooperative's board members. Most board members lack the resources or expertise to manage properties and their energy consumption. This ignorance presents a unique challenge, especially given the EU directives that prohibit buildings rated as energy classes F and G by 2033. 
Conversational agents (CAs) enable human-like interactions with computer systems, facilitating human-computer interaction across various domains. In our case, CAs can be implemented to support cooperative members in making informed energy retrofitting and usage decisions. 
This paper introduces a Conversational agent system, called SPARA, designed to advise cooperatives on energy efficiency. SPARA functions as an energy efficiency advisor by leveraging the Retrieval-Augmented Generation (RAG) framework with a Language Model(LM). The LM generates targeted recommendations based on a knowledge base composed of email communications between professional energy advisors and cooperatives' representatives in Stockholm. The preliminary results indicate that SPARA can provide energy efficiency advice with precision 80\%, comparable to that of municipal energy efficiency (EE) experts. A pilot implementation is currently underway, where municipal EE experts are evaluating SPARA performance based on questions posed to EE experts by BRF members. Our findings suggest that LMs can significantly improve outreach by supporting stakeholders in their energy transition. For future work, more research is needed to evaluate this technology, particularly limitations to the stability and trustworthiness of its energy efficiency advice.
\end{abstract}
\section{Introduction}
In Sweden, a substantial proportion of multifamily buildings are owned and managed by tenant-owner housing cooperatives (\textit{Swedish:} BRF, bostadsrättsförening). This ownership model provides its residents an autonomy in managing the co-owned property while ensuring compliance with energy use guidelines. Often, cooperative members lack the technical expertise required in energy efficiency and building maintenance, making it difficult to ensure compliance with the required guidelines. The challenge of improving building energy efficiency is exacerbated by the ageing of the building stock, which predates the prioritisation of energy efficiency standards \cite{IONESCU2015243}. Addressing the challenges of ageing building stock along with changing energy guidelines is essential, particularly in the context of Sweden's commitment to achieving both national and European Union (EU) climate goals.

The 2024 recast of the Energy Performance of Buildings Directive (EPBD) \cite{epbd_recast_2024} introduces stricter energy efficiency standards, requiring zero emissions for new buildings by 2030 and minimum energy class F by 2030 and E by 2033 for existing residential buildings. These regulations pose significant financial and technical challenges for BRFs in Sweden, many of which manage low-rated buildings and require extensive renovations to comply \cite{amoah_barriers_2022}.

Traditional energy efficiency guidance relies on costly consultants and technical tools, often inaccessible to BRFs. Many lack the financial resources and expertise to implement necessary retrofits. The available information, such as energy performance certificates and government guidelines, is  fragmented, overly technical, and difficult for non-experts to interpret. Uncertainty about data accuracy further discourages action, delaying regulatory compliance and sustainability progress. Addressing these barriers requires clear, reliable, and accessible guidance, tailored to non-expert decision-makers, enabling BRF boards to make informed and cost-effective energy improvements.

Conversational Agents (CA)\cite{allouch2021conversational}, or virtual assistants, are software systems designed to interact with users through Natural Language Processing (NLP) systems. These agents often replicate human-like behaviour and can be trained to generate answers to specific scenarios. In the context of energy efficiency, conversational agents can democratize access to expert knowledge. For BRFs, the agents can function as virtual energy advisors, bridging technical expertise gaps and aiding decision-making. By integrating knowledge from energy advisors, data from urban energy datasets, and EU regulations, these agents can provide tailored advice on energy retrofitting, energy class requirements, cost-effective renovation strategies, and compliance with EU regulations. Such decision-support tools can reduce the reliance on external consultants by offering continuous support and empowering BRF board members to make informed decisions. 

This paper presents SPARA, a system with conversational agents specifically developed to support BRFs in addressing the complexities of energy efficiency improvements and regulatory compliance. SPARA leverages Generative Pre-Trained Transformer (GPT-4o) as its core Language Model (LM), to function as a virtual energy advisor. SPARA's knowledge base combines insights from email exchanges between professional energy advisors and BRFs in Stockholm with data about energy usage of specific buildings. The integration of these diverse sources enables SPARA to bridge the gap between technical expertise and practical decision-making, providing accessible and actionable advice to non-experts.

The primary objectives of SPARA are to provide customized recommendations for energy retrofitting, enhance the decision-making capabilities of the BRF board members, and support compliance with EU energy regulations. The obtained results demonstrate SPARA's ability to correctly advise users about questions pertaining to energy efficiency. Pilot testing with municipal energy experts (\textit{Swedish:} EKR, Energi- och klimatrådgivningen) and BRFs has demonstrated the ability of SPARA to improve stakeholder support in energy transitions. However, evaluations also underscore the importance of ensuring the reliability, trustworthiness, and contextual relevance of the system while addressing limitations in scenarios that require highly specialized or dynamic advice.

\section{Related Work}
Conversational agents, CA, \cite{allouch2021conversational}, along with advances in NLP, has emerged as a sophisticated tool for human-computer interaction, enabling intuitive and dynamic exchanges across various domains. At their core, these agents rely on LMs, which are algorithms trained on extensive text datasets to predict and generate human-like language. Language models, LMs, such as GPT-3\cite{brown2020language}, extend this capability by incorporating 173 billion parameters, allowing them to process complex queries, generate contextually relevant responses, and integrate diverse data sources.

Retrieval-augmented generation (RAG) is a framework that improve the functionality of LM-based CA systems. RAG combines the generative capabilities of LMs with the precision of retrieval-based systems, allowing agents to incorporate domain-specific knowledge into their responses \cite{lewis2020retrieval}. In this RAG framework, CA retrieves relevant information from a predefined knowledge base, such as documents, datasets, or email communications, and integrates this information into its generated responses. This approach ensures that agent advice is contextually relevant and grounded in accurate and current information, making RAG particularly effective for applications, such as energy efficiency guidance, where domain-specific expertise is critical. The LM's to access external knowledge bases during interactions, ensures that their responses are grounded in relevant and up-to-date information. However, despite their capabilities, LMs are prone to \textit{hallucinations} \cite{10.1145/3571730}, where they generate factually inaccurate or misleading information with high confidence. Although model parameters, such as \textit{ temperature} \cite{franceschelli2024creativity} can be adjusted to control LM creativity, it does not always ensure consistent outputs, raising concerns about their \textit{stability} and \textit{reliability} \cite{liu2024trustworthyllmssurveyguideline}. These constraints highlight the importance of integrating robust retrieval systems, comprehensive evaluation metrics, and safeguards to ensure the trustworthiness of conversational agents in critical applications.  


SPARA, the conversational agent presented in this paper, leverages a LM to generate human-like text. To ensure contextually relevant energy efficiency advice, we employ retrieval-augmented generation (RAG) for traceability and set a low \textit{temperature} to minimize excessive creativity. Building on prior work in conversational agents \cite{allouch2021conversational} and RAG \cite{lewis2020retrieval}, SPARA addresses challenges like \textit{hallucinations} \cite{10.1145/3571730}, \textit{stability}, and \textit{reliability} \cite{liu2024trustworthyllmssurveyguideline}, enhancing precision by integrating expert knowledge from energy advisors.


\section{Conversational Agent System - SPARA}
A conversational agent system called SPARA (Smart Property Advisor for Retrofitting Actions) is a proposed solution designed to assist BRF members by addressing inquiries related to energy use and efficiency. In addition, SPARA supports informed decision making regarding energy retrofitting. The system is a CA that leverages LMs, specifically GPT-4o, to generate human-like responses and provide customized guidance to BRF members. To ensure that the model delivers contextually relevant advice tailored to buildings in Sweden, we integrate a dedicated knowledge base that provides additional contextual information. SPARA is built using an RAG architecture, which improves its ability to provide accurate and reliable responses. SPARA combines general language understanding with access to trusted building-related information to generate more precise and grounded answers.

\begin{figure}[h] 
    \centering
    \includegraphics[width=1\textwidth]{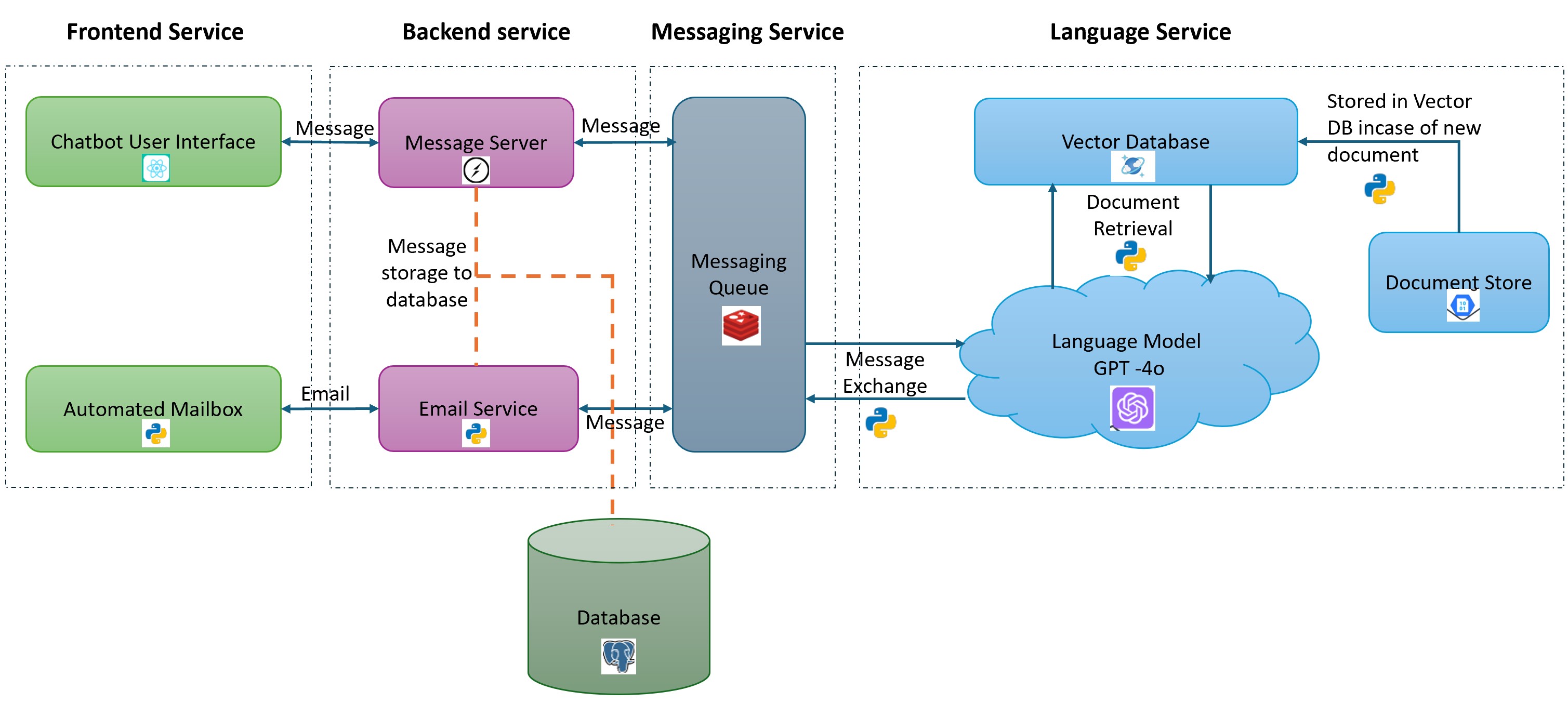} 
    \caption{Architecture of SPARA conversational agent system}
    \label{fig:spara_diagram} 
\end{figure}

The conversational agent (CA) system consists of four interconnected services, each responsible for specific functionalities. The architecture diagram above (Figure \ref{fig:spara_diagram}) provides a detailed overview of the components within these services.

\begin{enumerate}

\item{The \textbf{Frontend service} provides BRF members with multiple ways to interact with SPARA. This includes a user interface (UI) that functions as a chatbot, allowing users to pose queries and receive responses. The chatbot UI is developed using React. Additionally, users can contact SPARA by sending queries to a designated email address, enabling asynchronous interaction.}

\item{The \textbf{Backend service} facilitates communication between the front-end and the LM. WebSocket connections, implemented using Socket.IO, enable real-time interactions between the chatbot UI and the LM. When a user submits a query via email, a Python script processes the incoming message, extracts the relevant question, and forwards it to the system. The backend also includes a database that stores previous conversations along with the necessary user information, ensuring compliance with GDPR regulations. Messages are saved after a period of inactivity to support future context-aware interactions. This part acts as the system's internal coordinator — handling message exchange between the frontend and the language model and saving data securely and responsibly.}

\item{The \textbf{Messaging service} handles the queuing and distribution of user queries by utilizing a Redis-based message broker, which enables efficient and low-latency communication between system components \cite{zhang2015reliable}. Incoming queries are inserted into a Redis queue and subsequently processed by dedicated worker processes. In configurations where a single worker is employed, queries are processed sequentially, preserving their order. When multiple workers are active, Redis distributes queries across them, allowing for parallel processing but without guaranteed ordering. Upon completion, each response is also stored in Redis, ensuring traceability between queries and their corresponding answers. This architecture accommodates multiple simultaneous user interactions without overloading the system, while maintaining a reliable and transparent communication flow.}

\item{Finally, the \textbf{Language service}, which comprises the LM used for response generation - in this case, GPT-4o. Since SPARA employs an RAG approach, a dedicated knowledge base is integrated to improve response accuracy. This knowledge base contains building-related information, datasets, and relevant documents, all stored in Azure Blob Storage. Another essential component of the RAG architecture is a vector database, which stores high-dimensional embeddings. For this purpose, we utilize Azure Cosmos DB. This is the part of SPARA that understands questions and generates answers - enhanced with access to trusted energy and building information to ensure that the advice is relevant and accurate.}

\end{enumerate}

\begin{figure}[h] 
    \centering
    \includegraphics[width=1\textwidth]{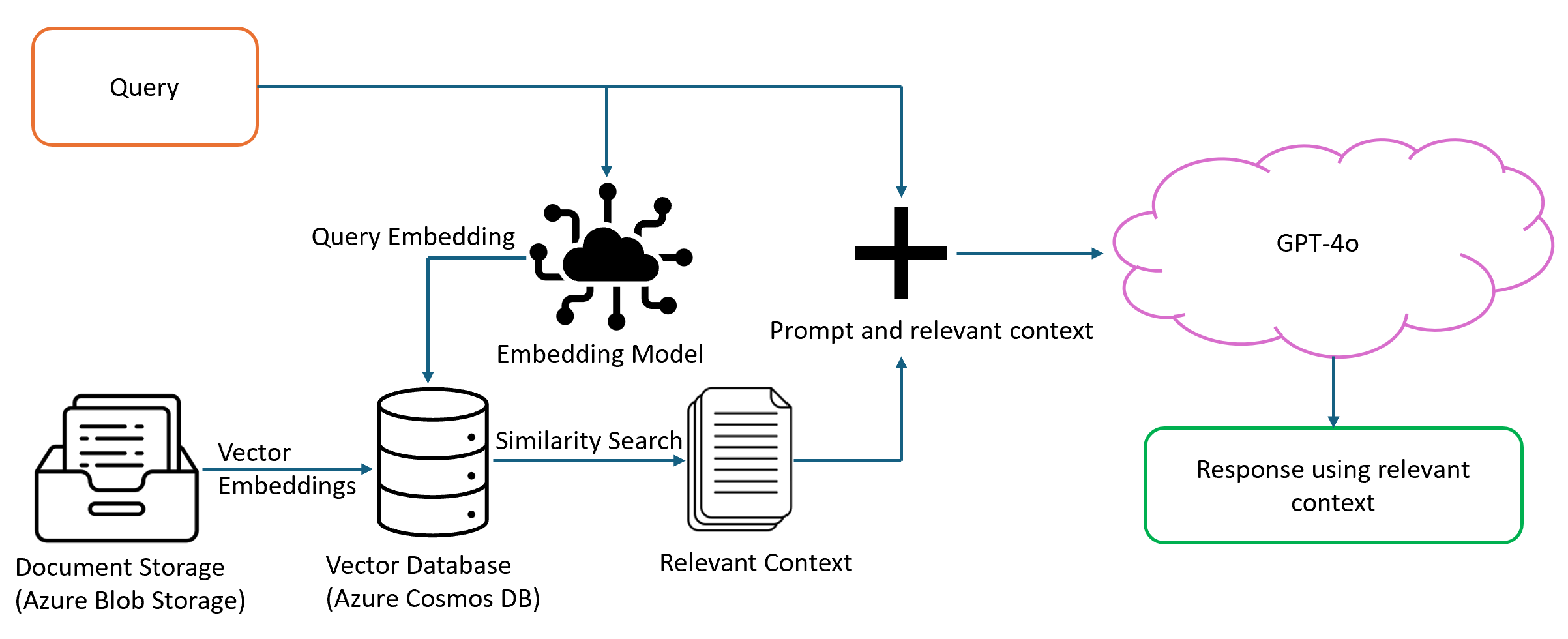} 
    \caption{Retrieval-augmentation generation (RAG) of response to a user query}
    \label{fig:rag_diagram} 
\end{figure}

RAG is an advanced framework that enhances the capabilities of LMs by combining generative modeling with retrieval techniques. Although LMs excel at generating coherent and contextually relevant text, they are often limited by their training data, which may not include the latest information or domain-specific knowledge. RAG addresses this limitation by incorporating external knowledge bases, enabling the model to retrieve and utilize relevant information during the query-response process. SPARA can refer to reliable documents and databases while responding to questions, rather than relying only on what it has 'learned' before. 

A \textbf{Knowledge Base} in this context refers to a structured or unstructured repository of information that contains domain-specific data, documents or facts. Examples include databases of EPCs, regulatory documents, technical manuals, and urban energy data sets. The knowledge base serves as a supplemental resource that the retrieval mechanism can query to provide accurate and up-to-date information, ensuring that the responses generated by the LM are grounded in factual and relevant data. As shown in the Figure\,\ref{fig:rag_diagram} above, we store the above documents in Azure Blob Storage.

\textbf{Retrieval techniques} are critical to the RAG framework, as they enable efficient extraction of relevant information from the knowledge base. These techniques often rely on methods such as semantic search, keyword matching, or vector-based similarity scoring. Semantic search, for example, involves encoding both the query and the knowledge base contents into high-dimensional vectors using embeddings generated by language models. The system then retrieves the most relevant documents or passages by calculating the similarity between these vectors. This enables SPARA to find specific, relevant content, even if the user's question is not exactly phrased like the document wording.

Once the retrieval system extracts relevant information, it passes the retrieved data as \textbf{context} to the LM input prompt. This process augments the model's understanding of the query by providing it with additional background knowledge. The input prompt, therefore, consists of the original user query, supplemented by the retrieved context. For example, in an energy efficiency application, the retrieved context might include specific EPC details, relevant EU regulations, or prior correspondence with energy advisors. The response is then shaped both by the user’s question and by real, up-to-date data relevant to that case.

The LM then processes the augmented prompt to generate a response. By combining the generative capabilities of the LM with the domain-specific precision of the retrieval system, RAG ensures that the output is contextually appropriate and factually accurate. This approach significantly improves the reliability and usefulness of conversational agents, especially in scenarios where accurate domain-specific knowledge is critical. In summary, RAG integrates the strengths of LMs and retrieval systems to create a powerful hybrid framework. Using external knowledge bases, retrieval techniques, and contextual augmentation, it ensures that the responses generated are grounded in real-world data while maintaining the conversational fluency of LM. This makes RAG particularly effective for applications like SPARA, where users require precise, actionable, and contextually relevant guidance for retrofit planning, where the guidance must be grounded in regulations, building characteristics, and energy performance metrics.

The diverse sources of knowledge, such as compliance data for EU energy regulations, metrics on energy consumption patterns for buildings, enables SPARA to address both the practical and regulatory dimensions of energy efficiency. The knowledge base not only enhances the accuracy of SPARA's responses but also ensures its advice remains grounded in current and contextually relevant information, thereby bridging the gap between technical expertise and actionable decision-making.

\section{Results}
The evaluation of text generated by LMs is a complex and multidimensional task, which requires assessing both semantic accuracy and contextual relevance. In this study, the evaluation of SPARA's responses was implemented in two parts: 1) classical NLP evaluation metrics, and 2) an analysis based on the nature of the question sets. This multifold approach enabled a more comprehensive assessment of the system's ability to provide energy efficiency advice. To assess the quality of the generated responses, we employed two widely adopted metrics: \textbf{cosine similarity} and \textbf{Jaccard similarity}. Cosine similarity quantifies the semantic similarity between two sentences by measuring the cosine of the angle between two vectors in a embedding space, effectively capturing the conceptual closeness between texts. . Jaccard similarity, on the other hand, focuses on lexical overlap by calculating the ratio of the intersection to the union of word sets, offering the insight into shared vocabulary and key terms.

The practical relevance of these metrics is best illustrated through examples. Consider a case where SPARA is asked how to reduce heat loss in a 1960s concrete apartment block. If the municipal energy experts recommend external wall insulation and window replacement, and SPARA suggests adding façade insulation and upgrading to energy-efficient glazing, cosine similarity would capture the shared technical meaning despite differing phrasing. Meanwhile, Jaccard similarity would reflect overlap in key terms such as "insulation", "windows", or "energy efficient", highlighting whether key concepts are present. In another case, SPARA is asked whether a building meets Class B energy requirements. If the system responds by reporting the building's current energy use and comparing it to the class B threshold, high similarity scores would indicate both  accurate interpretation of the question and a technically relevant response. Together, these metrics offer complementary perspectives: cosine similarity assesses semantic alignment, while Jaccard similarity reveals lexical consistency. When interpreted jointly, they provide a robust measure of whether SPARA’s advice meaningfully aligns with expert expectations in the context of building energy performance and retrofit planning.

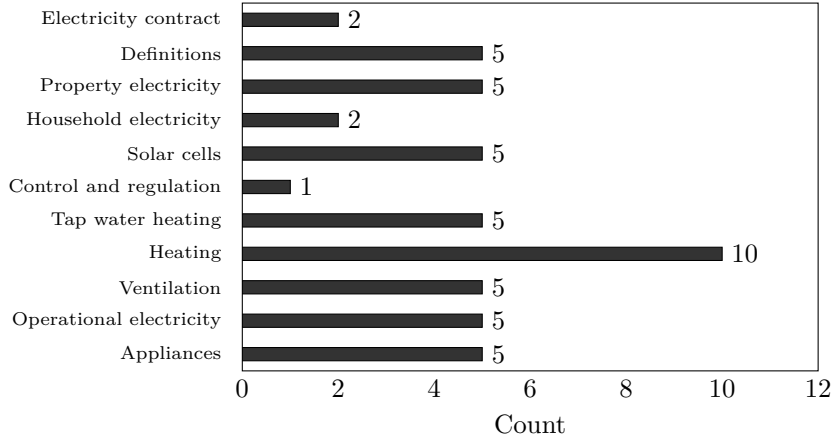
\begin{figure}[h]
    \centering
    \begin{tikzpicture}
        \begin{axis}[
            xbar,
            xmin=0, xmax=12,  
            symbolic y coords={
                Appliances, 
                Operational electricity, 
                Ventilation, 
                Heating, 
                Tap water heating, 
                Control and regulation, 
                Solar cells, 
                Household electricity, 
                Property electricity, 
                Definitions, 
                Electricity contract},
            ytick=data,
            nodes near coords,
            bar width=5pt,  
            xlabel={Count},
            yticklabel style={font=\scriptsize}, 
            xtick style={draw=none},  
            ytick style={draw=none},  
            enlarge y limits=0.05,  
            width=10cm,  
            height=7cm,  
            scale=0.9  
        ]
        \addplot [fill={rgb,1:red,0.2; green,0.2; blue,0.2}] coordinates {
            (5,Appliances) 
            (5,Operational electricity) 
            (5,Ventilation) 
            (10,Heating) 
            (5,Tap water heating) 
            (1,Control and regulation) 
            (5,Solar cells) 
            (2,Household electricity) 
            (5,Property electricity) 
            (5,Definitions) 
            (2,Electricity contract)
        };
        \end{axis}
    \end{tikzpicture}
    \caption{Distribution of evaluated questions \textit{(N = 50)} by topic categories}
    \label{fig:category_distribution}
\end{figure}

Given SPARA’s dual functionality in answering both general and building-specific questions about energy efficiency, the evaluation was conducted using two distinct question sets. The first set, referred to as general questions, addressed broad topics such as heating systems, ventilation, and insulation of outer walls. The answers to these questions were primarily text-based and descriptive. The second set, termed specific questions, focused on building-level data, including queries about district heating requirements and total energy consumption. The responses in this category were predominantly numeric, reflecting SPARA's ability to process and generate data-driven output.

For text-based questions, the evaluation followed a systematic methodology. First, a control set of questions and their corresponding reference answers were created. These questions covered a wide range of topics related to energy efficiency to ensure robustness. The control questions were then posed to SPARA, and the generated responses were recorded. Both the reference answers and SPARA's generated responses were converted into vector embeddings using a pre-trained language model. This transformation enabled the application of cosine similarity to measure the semantic alignment between the reference text and the generated text. Jaccard similarity was also calculated to capture the lexical overlap, providing an additional perspective on the differences between the two sets of responses.

For building-specific questions, the evaluation focused on the accuracy of the numeric outputs. Each query was compared directly with the reference data from the control set. This ensured that SPARA’s ability to process and generate precise, data-driven recommendations could be quantitatively validated.

The control set for text-based questions was constructed by sampling inquiries in multiple categories, including \textbf{building heating}, \textbf{hot water circulation}, and \textbf{solar cell utilization}. These categories were selected to ensure a comprehensive representation of topics related to \textbf{energy efficiency}. In total, \textbf{50} distinct questions were chosen for evaluation. 

These questions were processed using the \textbf{SPARA} system, which generated the corresponding responses stored in the \textit{generated response} column, while the reference answers were maintained in the \textit{answer} column. To quantify the similarity between the generated and actual responses, \textbf{Jaccard similarity} and \textbf{cosine similarity} were employed. Cosine similarity was computed using \textbf{GPT-based embeddings}, leveraging high-dimensional vector representations to capture semantic similarity, while Jaccard similarity was calculated based on \textbf{word-level overlap}.  

\begin{figure}[h]
    \centering
    \begin{tikzpicture}
        \begin{axis}[
            ybar,
            symbolic x coords={0-0.2, 0.2-0.4, 0.4-0.6, 0.6-0.8, 0.8-1.0},
            xtick=data,
            ymin=0, ymax=55, 
            ylabel={Count},
            xlabel={Similarity Range},
            legend pos=north west, 
            bar width=6pt, 
            nodes near coords, 
            enlarge x limits=0.1, 
            width=10cm, 
            height=6cm, 
            scale=0.9 
        ]
        
            \addplot coordinates {(0-0.2,4) (0.2-0.4,2) (0.4-0.6,3) (0.6-0.8,0) (0.8-1.0,41)};
            \addlegendentry{Jaccard Count}

            \addplot coordinates {(0-0.2,0) (0.2-0.4,1) (0.4-0.6,0) (0.6-0.8,2) (0.8-1.0,47)};
            \addlegendentry{Cosine Count}

        \end{axis}
    \end{tikzpicture}
    \caption{Level of lexical and semantic alignment between tested questions \textit{(N = 50)} and SPARA responses evaluated by Jaccard \textit{(blue)} and cosine \textit{(red)} similarity scores respectively}
    \label{fig:similarity_comparison}
\end{figure}
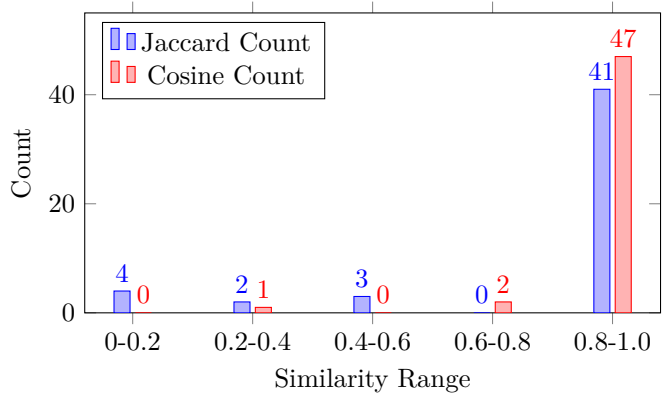

The distribution of similarity scores indicates a strong agreement between the generated and reference responses. In terms of \textbf{Jaccard similarity}, \textbf{41 out of 50 responses} achieved a similarity score above \textbf{0.8}, suggesting a high degree of lexical overlap, while \textbf{4 responses} had scores below \textbf{0.2}, indicating notable discrepancies. For \textbf{cosine similarity}, an even higher proportion of responses (\textbf{47 out of 50}) fell within the \textbf{0.8–1.0 range}, reinforcing the strong \textbf{semantic alignment} captured by embeddings.  

These findings suggest that the majority of responses generated by \textbf{SPARA} exhibit \textbf{high lexical and semantic similarity} to the reference answers, demonstrating the system’s ability to produce responses that closely align with the expected outputs.  

The results for building-specific questions were equally promising as those for text-based questions. SPARA was provided with information about approximately ten buildings, covering various aspects such as the declared energy class of the buildings and the consumption of electricity for different uses, including hot water, laundry rooms, district heating, and others. Ten such questions were posed to SPARA, and the responses were recorded. SPARA correctly answered 85\% of the total questions asked. Furthermore, SPARA was able to provide a detailed breakdown of energy usage for a specific month of the year, as well as detailed information by building. However, it did not provide accurate answers when the questions required SPARA to compute electricity data over multiple months in a given year.

\begin{table}[h]
    \centering
    \caption{Sample Questions and Answers with Acceptance Status}
    \label{tab:qa_acceptance}
    \renewcommand{\arraystretch}{1} 
    \setlength{\tabcolsep}{8pt} 
    \small
    \begin{tabular}{|p{6cm}|p{6cm}|p{1.8cm}|}
        \hline
        \textbf{Question} & \textbf{Answer} & \textbf{Acceptance} \\
        \hline
        What is the normal household eui for building id 5? &
        The normal household energy use intensity (EUI) for building id 5 is 30.00 kWh/m². & Correct \\
        \hline
        What is the deduction in household heat electricity for building id 11? &
        The deduction in household heat  electricity for building id 11 is 41755.50. & Correct \\
        \hline
        What is the total electricity use in laundry room for building id 5 for the month of August for every year? Give a detailed breakdown. &
        I'm sorry, but the context provided does not contain information about the total electricity use in the laundry room for building id 5 for the month of August for every year. & Incorrect \\
        \hline
    \end{tabular}
\end{table}
Currently, in the validation stage SPARA is evaluated by EKR experts to determine its ability to generate accurate and contextually relevant advice on building retrofitting and energy efficiency. The evaluation involves a series of interactions in which experts pose a diverse set of queries to SPARA, covering both general topics in energy efficiency and specific building-level requests. Communication between EKR experts and SPARA is facilitated through an automated email-based workflow. Emails containing queries are forwarded to a dedicated SPARA email account, where they are processed by SPARA’s email service. The service analyzes the content of the incoming emails, generates context-specific responses using the RAG framework, and returns the responses to the EKR experts. The experts then rate these responses to evaluate their quality, specifically assessing the precision, relevance, and overall usefulness of the advice provided. This setup enables a systematic assessment of SPARA performance in a real-world context while maintaining a standardized mechanism to collect expert feedback.

\section{Conclusions}
This paper introduces SPARA, a conversational agent system designed to assist with energy-related inquiries. The system provides energy efficiency advice and retrofit recommendations for buildings. Currently, SPARA is being piloted for testing with energy experts from the City of Stockholm, as well as representatives from eight housing cooperatives. The preliminary results of these tests indicate that SPARA effectively responds to general inquiries about energy usage. However, when faced with domain-specific or ambiguous questions that require expert knowledge, SPARA refrains from providing speculative responses. Ensuring that the system does not generate speculative answers was a key performance metric, and the results demonstrate that this objective has been successfully achieved.

SPARA is currently being used by housing cooperatives to better understand their current energy performance, interpreting measured energy usage and performance metrics, as well as to identify feasible retrofitting measures and assess their expect impact for energy saving and emission reduction. The system offers data-driven suggestions based on available energy declarations, enabling cooperatives to estimate potential energy savings and financial implications of retrofitting actions. It also supports benchmarking functionality, which enables users to compare the energy performance of their building with that of similar buildings in the region. Municipal energy experts use SPARA to quickly retrieve building-specific information, reducing the manual effort typically involved in accessing and interpreting building data. While the system is still in the limited testing phase, the early reports by experts indicate that SPARA performs satisfactorily when tested against a variety of use cases. Hence, SPARA bridges a critical gap between technical information such as building codes, expert handbooks and building-specific energy performance data and practical decision making. Its further integration into planning processes can streamline the early stages of retrofit planning, providing stakeholders with initial guidance for commissioning of energy audits and feasibility studies, as well as later retrofits, including their monitoring and validation.

\section{Acknowledgments}
This study is part of the projects SPARA (P2022-00925) and DigiCityClimate financed by the Swedish Energy Agency via the E2B2 programme and Google via ICLEI Action Fund 2.0, respectively. The authors express gratitude to our MSc students Daniel Nuñez Hernandez and Zelal Irem Aldag Bozkurt for their great contribution in the practical development of SPARA.

\bibliographystyle{vancouver}
\bibliography{references} 

\begin{thebibliography}{10}

\bibitem{IONESCU2015243}
Ionescu C, Baracu T, Vlad GE, Necula H, Badea A.
\newblock The historical evolution of the energy efficient buildings.
\newblock Renewable and Sustainable Energy Reviews. 2015;49:243-53.
\newblock Available from: \url{https://www.sciencedirect.com/science/article/pii/S1364032115003329}.

\bibitem{epbd_recast_2024}
{European Parliament}. Directive ({EU}) 2024/1275 of the {European} {Parliament} and of the {Council} of 24 {April} 2024 on the energy performance of buildings (recast) ({Text} with {EEA} relevance); 2024.
\newblock Accessed: 2025-02-02.
\newblock Available from: \url{http://data.europa.eu/eli/dir/2024/1275/oj/eng}.

\bibitem{amoah_barriers_2022}
Amoah C, Smith J.
\newblock Barriers to the green retrofitting of existing residential buildings.
\newblock Journal of Facilities Management. 2022 May;22(2):194-209.
\newblock Publisher: Emerald Publishing Limited.
\newblock Available from: \url{https://www.emerald.com/insight/content/doi/10.1108/jfm-12-2021-0155/full/html}.

\bibitem{allouch2021conversational}
Allouch M, Azaria A, Azoulay R.
\newblock Conversational agents: Goals, technologies, vision and challenges.
\newblock Sensors. 2021;21(24):8448.

\bibitem{brown2020language}
Brown T, Mann B, Ryder N, Subbiah M, Kaplan JD, Dhariwal P, et~al.
\newblock Language models are few-shot learners.
\newblock Advances in neural information processing systems. 2020;33:1877-901.

\bibitem{lewis2020retrieval}
Lewis P, Perez E, Piktus A, Petroni F, Karpukhin V, Goyal N, et~al.
\newblock Retrieval-augmented generation for knowledge-intensive nlp tasks.
\newblock Advances in neural information processing systems. 2020;33:9459-74.

\bibitem{10.1145/3571730}
Ji Z, Lee N, Frieske R, Yu T, Su D, Xu Y, et~al.
\newblock Survey of Hallucination in Natural Language Generation.
\newblock ACM Comput Surv. 2023 Mar;55(12).
\newblock Available from: \url{https://doi.org/10.1145/3571730}.

\bibitem{franceschelli2024creativity}
Franceschelli G, Musolesi M.
\newblock On the creativity of large language models.
\newblock AI \& SOCIETY. 2024:1-11.

\bibitem{liu2024trustworthyllmssurveyguideline}
Liu Y, Yao Y, Ton JF, Zhang X, Guo R, Cheng H, et~al.. Trustworthy LLMs: a Survey and Guideline for Evaluating Large Language Models' Alignment; 2024.
\newblock Available from: \url{https://arxiv.org/abs/2308.05374}.

\bibitem{zhang2015reliable}
Zhang T.
\newblock Reliable event messaging in big data enterprises: looking for the balance between producers and consumers.
\newblock In: Proceedings of the 9th ACM International Conference on Distributed Event-Based Systems; 2015. p. 226-33.

\end{thebibliography}

\end{document}